\documentclass[final,3p,times]{elsarticle}
\pdfoutput=1

\usepackage{graphicx}
\DeclareGraphicsExtensions{.pdf,.jpg,.jpeg}
\usepackage[cmyk]{xcolor}
\usepackage{cancel}
\usepackage{eso-pic} 
\usepackage{comment} 
\usepackage{amsmath}
\usepackage{breqn}

\journal{Journal of Computational Physics}

\begin{document}

\begin{frontmatter}



\title{Novel Kinetic 3D MHD Algorithm \\
for High Performance Parallel Computing Systems }


\author[kiam]{B.~Chetverushkin}
\author[desy]{N.~D'Ascenzo}
\author[kiam,desy]{V.~Saveliev \corref{cor1}}
\ead{saveliev@mail.desy.de}
\cortext [cor1]{Corresponding author }
\address[kiam]{Keldysh Institute of Applied Mathematics, Russian Academy of Science, Russia}
\address[desy]{Deutsches Elektronen Synchrotron, Germany}

\begin{abstract}
The impressive progress of the kinetic schemes in the solution of gas dynamics problems and the development of effective parallel algorithms for modern high performance parallel computing systems led to the development of advanced methods for the solution of the magnetohydrodynamics problem in the important area of plasma physics.
The novel feature of the method is the formulation of the complex Boltzmann-like distribution function of kinetic method with the implementation of electromagnetic interaction terms.  
The numerical method is based on the explicit schemes. Due to logical simplicity and its efficiency, the algorithm is easily adapted to modern high performance parallel computer systems including hybrid computing systems with graphic processors.
\end{abstract}

\begin{keyword}
Magnetohydrodynamics (MHD) \sep kinetic scheme \sep high performance computing
\end{keyword}
\end{frontmatter}


\section{Introduction}
\label{introduction}
The tremendous progress in the development of high performance computing systems, especially expecting drastically new exascale computing systems, including the challenges in architecture, scale, power and reliability, gives new opportunities for the mathematical modeling of important physical phenomena in the present and future. Nevertheless the complexity of the challenges in science and engineering continues to outpace our ability to adequately address them through impressively growing computational power.

A feature of the present is that the development of technologies and computer systems architecture are well ahead of software development. The software problems are primarily associated with the complexity of the algorithms adaptation for the differential equations of mathematical physics to high performance computing systems architecture.
In particular they refer to one of the important requirements as the accuracy in combination with the correctness of the initial mathematical models. Another requirement for the methods is their logical simplicity and high efficiency at the same time. The numerical algorithms should be simple and transparent from a logical point of view.   

One of the important directions to overcome these problems is the development of a nontraditional approach to initial mathematical models and computational algorithms. In the present study for the solution of the multidimensional gas dynamics and magnetohydrodynamics problems kinetic difference scheme is proposed. It is convenient from the physics point of view to define the gas dynamics and magnetohydrodynamics quantities from close relations between the kinetic and gas dynamics description of physics processes \cite{Chetver_1,Chetver_2}.
  
Another aspect is the study of the explicit finite difference schemes, which seem to be preferable for future high performance parallel computing, especially in terms of their simplicity and well adaptability to parallel program realization, including hybrid high performance parallel computing systems. The weakness of explicit schemes is a strictly limited time step that ensures computational stability. This restriction becomes critical with the growing number of nodes and the reduction in the step of a spatial mesh. The advanced explicit kinetic finite difference schemes have a soft stability condition giving the opportunity to enhance the stability and to use very fine meshes~\cite{Chetver_3}.

The mentioned aspects are used for the development of the framework for the study of the dynamics of the conducting gas media in strong magnetic fields at high performance parallel computing systems. 

\section{Theoretical Issues}
\subsection{Gas Dynamics Processes}
The kinetic theory describes the gas dynamics by the Boltzmann differential equation through the evolution of the distribution function $f\left(\mathbf{x},\boldsymbol\xi,t\right)$~\cite{Maxwell_1}:
\begin{dmath}
\frac{\partial f\left(\mathbf{x},\boldsymbol\xi,t\right)}{\partial t}+\boldsymbol\xi\cdot\nabla f\left(\mathbf{x},\boldsymbol\xi,t\right)=C\left(f\right)
\label{boltzmanneq}
\end{dmath}
\hspace{1.0 cm}where $C\left(f\right)$ is a nonlinear integral operator which describes the collisions between gas molecules. 

\vspace*{0.25cm}
This evolution equation follows naturally from the relations between the kinetic and the gas dynamics description of continuous media. The macroscopic observables such as density, momentum, energy flux as a function of $\mathbf x$ and $t$ are obtained from the moments of the distribution function with respect to the macroscopic velocity. The evolution equations for these gas dynamics quantities are obtained by integrating  Eq.~(\ref{boltzmanneq}) over molecular velocities $\boldsymbol\xi$ with summational invariants $m$,$m\boldsymbol\xi$,$\frac{1}{2}m\boldsymbol\xi^{2}$.  
The computational interest in kinetic formulations of the gas dynamics is high due to the linearity of the differential operator on the left side of Eq.~(\ref{boltzmanneq}). Nonlinearity is confined by the collision term, which is generally local in $\mathbf x$ and $t$. 

An important feature is that the collision integral vanishes in the equilibrium state when the local Boltzmann distribution function $f$ is a Maxwellian:   

\begin{dmath}[compact]
f(\mathbf x, \boldsymbol \xi , t) = \frac{\rho(\mathbf x,t) m^{1/2}}{(2\pi kT\left (\mathbf x,t)\right )^{3/2} } 
exp \left \{ -\frac{m}{2kT(t,\mathbf x) }     (\boldsymbol \xi -\mathbf u(t,\mathbf x))^2  \right \}
\end{dmath}

This leads to the use of this model for numerical methods and possible generalizations in order provide a natural kinetic description of the system of conservation laws. This approximation is sufficient for the gas dynamics processes and is called the kinetic approach~\cite{Chetver_1}. 

\subsection{Electromagnetic Processes}
In~\cite{Tonks_1} it was shown that electromagnetic fields do not destroy the validity of the Boltzmann equation and this opened the way for the implementation of the electromagnetic term in the Boltzmann-like distribution function. From the vector nature of the electromagnetic interaction, the distribution function should taking to account the vector behavior and  provide correct formulation for the evolution of the magnetic field, i. e. the magnetic field should be generally defined as the momentum of the Boltzmann-like distribution function.

A few useful attempts to formulate the vector Boltzmann-like distribution function can be found in  ~\cite{Croisille_1, Huba_1, Dellar_1}, but physical meaning was not clear defined.

We propose an evaluation of the electromagnetic processes in the context of the distribution function, taking to account the axial nature of the magnetic field. The electromagnetic field is considered as a complex vector field as proposed in ~\cite{Landau_1}: 

\begin{dmath}[compact]
\mathbf F=\mathbf{E}+i\mathbf{B}
\end{dmath}

For the purposes of magnetohydrodynamics, the effect, which a magnetic field exerts on a certain volume, is obtained by integrating the electromagnetic stress tensor over the surface of that volume and the correspondent propagation velocity can be defined as a complex vector of velocity: 

\begin{dmath}[compact]
\mathbf v_{em}=\mathbf{u}_{em}+i\mathbf w_{em}
\end{dmath}

At first approximation the term defined of electric forces could be neglected and the magnetic term could be defined through the tension of the magnetic field line and shows a similarity to the Alfven wave mechanism:

\begin{dmath}[compact]
\mathbf w_{em}=\frac{\mathbf{B}}{\sqrt \rho}
\end{dmath}

\subsection{Proposed Distribution Function for MHD}
Using the above definitions we define the local complex Boltzmann Maxwellian distribution function of magnetohydrodynamics with drift velocity $\mathbf{u}$ in magnetic field $\mathbf B$ at the equilibrium:

\begin{dmath}[compact]
f_{M}\left(\mathbf{x},\boldsymbol\xi,t\right)
=\frac{\rho\left(\mathbf{x},t\right) m^{-1/2}}{\left(2\pi kT\left(\mathbf{x},t\right) \right)^{3/2}} 
\exp\left\{ {-\frac{m}{2kT\left(\mathbf{x},t\right)} \left|\left(\boldsymbol\xi-\mathbf{u}\left(\mathbf{x},t\right)\right) 
-i \mathbf w_{em}   \right|^{2}} \right\}
\label{fmhd}
\end{dmath}

The first term on the right-hand side of (\ref{fmhd}) includes the internal energy and the second term is the magnetic field energy.  The hydrodynamics observables are real scalars and vectors. The complex components include the dynamics of the macroscopic observables introduced by the evolution of the magnetic field, keeping their specific  pseudo-vectorial nature.

The magnetogasdynamics observables are obtained as integrals of the distribution function~(\ref{fmhd}) with the summational invariants $\left(m,m\boldsymbol\xi,\frac{1}{2}m\boldsymbol\xi^{2},m\boldsymbol\xi^{\ast}\right)$. The integration is performed on the path $\gamma$ with respect to the molecular velocities $\boldsymbol\xi$ in the complex plane correspondent to the value of $\mathbf B$ in the imaginary space. 
The relations are obtained for the real and imaginary terms:
\begin{dgroup}
\begin{dmath}[compact]
\rho\left(\mathbf{x},t\right)=\int_{\gamma}{m f_{M} d^{3}\boldsymbol\xi } 
\end{dmath}
\begin{dmath}[compact]
\mathbf{u}\left(\mathbf{x},t\right)=\frac{1}{\rho\left(\mathbf{x},t\right)}\int_{\gamma}{m \boldsymbol\xi f_{M} d^{3}\boldsymbol\xi}
\end{dmath}
\begin{dmath}[compact]
E\left(\mathbf{x},t\right)=\int_{\gamma}{\frac{1}{2}m \boldsymbol\xi^{2} f_{M} d^{3}\boldsymbol\xi}
\end{dmath}
\begin{dmath}[compact]
\mathbf{B}\left(\mathbf{x},t\right)=-\frac{1}{\sqrt{\rho\left(\mathbf{x},t\right)}}\int_{\gamma}{m \boldsymbol{\xi}^{\ast} f_{M} d^{3}\boldsymbol\xi }
\end{dmath}
\label{hydq}
\end{dgroup}

The proposed complex Boltzmann Maxwell like distribution function contains the hydrodynamics terms and the electromagnetic terms. Thus by using this distribution function to calculate the mass, momentum, energy and magnetic field fluxes, most of the electromagnetic contributions are calculated directly, i.e. one does not have to solve the hydrodynamics and magnetic force components separately or differently, as will be shown below.  

\section{Ideal MHD System of Equation }
To provide the first step of the formulation of the MHD conservation laws equation,
the equilibrium state is considered with the proposed distribution function. 
The MHD system of equations is obtained by the integration of (\ref{boltzmanneq}) with vanishing collision integral with the summational invariants following the definition in (\ref{hydq}):

\begin{dgroup}[noalign]
\begin{dmath}[compact]
\int_{\gamma}{m\frac{\partial f}{\partial t}}+\int_{\gamma}{m\boldsymbol\xi\cdot\nabla f d^{3}\boldsymbol\xi}=0
\end{dmath}
\begin{dmath}[compact]
\int_{\gamma}{m\boldsymbol\xi\frac{\partial f}{\partial t}}+\int_{\gamma}{m\boldsymbol\xi\boldsymbol\xi\cdot\nabla f d^{3}\boldsymbol\xi}=0
\end{dmath}
\begin{dmath}[compact]
\int_{\gamma}{\frac{1}{2}m\boldsymbol\xi^{2}\frac{\partial f}{\partial t}}+\int_{\gamma}{\frac{1}{2}m\boldsymbol\xi^{2} \boldsymbol\xi \cdot \nabla f d^{3}\boldsymbol\xi}=0
\end{dmath}
\begin{dmath}[compact]
\frac{1}{\sqrt{\rho}}\int_{\gamma}{m\boldsymbol\xi^{\ast}\frac{\partial f}{\partial t}}+\frac{1}{\sqrt{\rho}}\int_{\gamma}{m\boldsymbol\xi^{\ast}\boldsymbol\xi\cdot\nabla f d^{3}\boldsymbol\xi}=0
\end{dmath}
\label{mhd3d}
\end{dgroup}

The result obtained, set of Eq.~(\ref{mhd3d}), is the ideal magnetohydrodynamics system of equations: 
\begin{dgroup}[noalign]
\begin{dmath}[compact]
\frac{\partial \rho}{\partial t} + \frac{\partial}{\partial x_{i}}\rho u_{i}=0
\end{dmath}
\begin{dmath}[compact]
\frac{\partial}{\partial t}\rho u_{i} + \frac{\partial}{\partial x_{k}}\left[\left(p+\frac{ B^{2}} {2} \right)\delta_{ik}+\rho u_{i}u_{k}-B_{i}B_{k} \right] =0
\end{dmath}
\begin{dmath}[compact]
\frac{\partial E}{\partial t}+\frac{\partial}{\partial x_{i}}\left[u_{i}\left(E+p+\frac{B^2}{2}\right)-B_{i}{u_{k}B_{k}}\right]=0
\end{dmath}
\begin{dmath}[compact]
\frac{\partial B_{i}}{\partial t}+ \frac{\partial}{\partial x_{k}}\left[u_{k}B_{i}-u_{i}B_{k}\right]=0
\end{dmath}
\label{mhd3d}
\end{dgroup}

In addition an equation for $\nabla\cdot\mathbf{B}$ is obtained as the imaginary part of the path integral of the summational invariant $\left(m\right)$ with respect to the velocities $\boldsymbol\xi$:

\begin{dmath}[compact]
\frac{1}{\sqrt{\rho}}\int_{\gamma}{m\frac{\partial f}{\partial t}}+\frac{1}{\sqrt{\rho}}\int_{\gamma}{m\boldsymbol\xi\cdot\nabla f d^{3}\boldsymbol\xi}=0
\label{divb}
\end{dmath}

\begin{dmath}[compact]
\frac{\partial  B_{i}}{\partial x_{i}}=0
\label{divb}
\end{dmath}


\section{Kinetic MHD Finite Difference Scheme}
The model of the kinetic differential schemes is based on the discrete model of evolution of the distribution function. Kinetic schemes are obtained directly from the Boltzmann kinetic equation by using the principle of total approximation.

Consider the local volume of gas (cell $i$) with the distribution function in time $t^{j}$. By using the splitting method of particle flow for the cell $i$, the evolution of the distribution function by first order differential scheme for the kinetic Boltzmann equation can be written as:
\begin{dmath}
\frac{f^{j+1}-f^{j}}{\Delta t} =\xi \frac{f_{i+1}^j - f_{i-1}^j}{2\Delta x} - 
\frac{\left |\xi \right | }{2}\frac{f^{j}_{i+1}-2f^{j}_{i}+f^{j}_{i-1}}{\Delta x} + C(f^J)
\label{balance_sc1}
\end{dmath}

As mentioned before the collision of particles leads to the establishment of the equilibrium state which is adequately described by the single-particle Maxwell distribution function with vanishing of the collision integral in the right part of the balance relations.
The time evolution of the distribution function can be represented as the time evolution of the local Maxwellian distribution function in discrete moments:  
\begin{itemize}
\item at time $t^{j}$, on each cell, the locally constant one-particle Maxwellian distribution function is defined:

\begin{dmath}
f_{M}
=\frac{\rho m^{-1/2}}{\left(2\pi kT \right)^{3/2}} \exp\left\{ -\frac{m}{2kT} \left|\left(\boldsymbol\xi-\mathbf{u}\right) -i\frac{\mathbf{B}}{\sqrt{\rho }}\right|^{2}\right\}
\end{dmath}
\hspace{0.5cm} where the magnetohydrodynamics parameters $\rho$,$\mathbf u$,$T$,$\mathbf B$ are not varied on the cell.
\item during the time interval $\Delta t = t^{j+1}-t^{j}$ collisionless processes of the gas dynamics occurs,
\item at time $t^{j+1}$ the distribution function is instantaneously maxwellised 
\item for the time $t^{j+2}$ these processes are repeated.
\end{itemize}

The kinetic difference scheme in this case can be written:
\begin{dmath}
\frac{f^{j+1}_i-f^{j}_{i,M}}{\Delta t} =\xi \frac{f^j_{i+1} - f_{i-1}^j}{2\Delta x} - 
\frac{\Delta x \left |\xi \right | }{2}\frac{f^j_{i+1,M}-2f^j_{i,M}+f^j_{i-1,M}}{\Delta x^2}
\label{balance_sc2}
\end{dmath}
or in more general form for the multidimensional case:
\begin{dmath}
\frac{f^{j+1}-f_{M}^{j}}{\Delta t}+\frac{1}{\Delta V}\xi_{i}f_{\sigma}^{j}\Delta\sigma_{i}=\frac{1}{2\Delta V}\left|\xi_{i}\right|\Delta x_{i}\frac{\partial f^{j}}{\partial x_{i}} \Delta\sigma_{i}
\label{balance_sc3}
\end{dmath}

\hspace*{1.cm} where: \\
\hspace*{1.cm} $\Delta\sigma_{i}$ is the surface element $\Delta x_{k}\Delta x_{m}$ perpendicular to the direction $x_{i}$,\\ 
\hspace*{1.cm } $f_{\sigma}^{j}$ is the value of the distribution function at the surface $\sigma$ between the two volume elements $I_{i}$ and $I_{i+1}$,\\
\hspace*{1.cm} $\frac{\partial f^{j}}{\partial x_{i}}$ the distribution function derivative at the surface $\sigma$ between the two volume elements.

The sum in Eq.~(\ref{balance_sc3}) is extended to the 6 surface elements at the boundary of the 3-dimensional rectangular volume element. 

\vspace*{0.25cm}
The kinetic scheme of the conservation laws of the macroscopic observables for 3D magnetohydrodynamics processes can be obtained by integrating the balance relation (\ref{balance_sc3}) with the summational invariants $m,m\boldsymbol\xi,
\frac{1}{2}m\boldsymbol\xi^{2},m\boldsymbol\xi^{\ast}$, using the same integration rules as in Eq.~(\ref{mhd3d}):
\begin{dgroup}[noalign]
\begin{dmath}[compact]
\frac{\rho^{j+1}-\rho^{j}}{\Delta t}+\left(\rho u_{i}\right)_{\hat{x}_{i}}=\frac{\Delta x_{i}}{2}\left[\rho u_{i}\mathrm{Erf}\left(\beta u_{i}\right)+\frac{\rho}{\beta\sqrt{\pi}}e^{-\beta^2 u_{i}^{2}}\right]_{\bar{x}_{i}x_{i}}
\end{dmath}
\begin{dmath}[compact]
\frac{\rho^{j+1}u^{j+1}_{i}-\rho^{j}u^{j}_{i}}{\Delta t}+\left[\rho u_{i}u_{k}+ \left (p+\frac{B^2}{2} \right )\delta_{ik}-B_{i}B_{k} \right]_{\hat{x}_{k}}= \\
\hspace*{1.5cm}\frac{\Delta x_{k}}{2}\left[\frac{\rho u_{i}}{\sqrt{\pi}\beta}e^{-\beta^{2}u_{k}^{2}}+\left(\rho u_{i}u_{k}+\left(p+\frac{B^2}{2}\right)\delta_{ik}\right)\mathrm{Erf}\left(\beta u_{k} \right)-B_{i}B_{k}\right]_{\bar{x}_{k}x_{k}}
\end{dmath}
\begin{dmath}[compact]
\frac{E^{j+1}-E^{j}}{\Delta t}+\left[u_{i}\left(E+p+\frac{B^2}{2}\right)-B_{i}u_{k}B_{k}\right]_{\hat{x}_{i}} = \\
\hspace*{-4.2cm} \frac{\Delta x_{i}}{2}\left[u_{i}\left(E+p+\frac{B^2}{2}\right)\mathrm{Erf}\left(\beta u_{i} \right)+    \frac{E+\frac{1}{2}\left(p+\frac{B^2}{2} \right)}{\beta\sqrt{\pi}}e^{-\beta^{2}u_{i}^{2}} -B_{i}u_{k}B_{k}\right]_{\bar{x}_{i}x_{i}}
\end{dmath}
\begin{dmath}[compact]
\frac{B^{j+1}_{i}-B^{j}_{i}}{\Delta t}+\left(u_{k}B_{i}-u_{i}B_{k}\right)_{\hat{x}_{k}}
= \frac{\Delta x_{k}}{2} \left [\frac{ B_{i}} {\sqrt{\pi}\beta} e^{-\beta^{2}u_{k}^{2}}  + B_{i}u_{k} \mathrm{Erf} \left(\beta u_{k} \right) -u_{i} B_{k}\right ]_{\bar{x}_{k}x_{k}}
\end{dmath}
\label{MHDscheme}
\end{dgroup} 

\hspace*{1. cm} where $\beta=\sqrt{\frac{\rho}{2p+B^2}}$, $i,k=1\ldots 3$ 
\vspace*{0.25 cm}

In addition to Eq.~(\ref{MHDscheme}) the condition is obtained  as the complex part of  the path integral of  the summational invariants $\left(m\right)$ with respect to the molecular velocities $\boldsymbol\xi$ in the complex plane:
\begin{dmath}
\left( B_{i}\right)_{\hat{x}_{i}}=\frac{\Delta x_{i}}{2}\left[  B_i \right]_{\bar{x}_{i}x_{i}}
\end{dmath}

Dissipative terms appear in the time evolution of the magnetic field which does not preserve the condition $\nabla\cdot\mathbf{B}=0$ and require a specific treatment.

\section{Kinetic Quasi MHD Equations}
The kinetic quasi magnetohydrodynamics system of equations is closely related to the kinetic scheme and represents a differential form notation for the numerical algorithms.

The balance relation in Eq.~(\ref{balance_sc3}) can be rewritten as:
\begin{dmath}[compact]
\frac{\partial f}{\partial t}+\frac{1}{\Delta V}\int_{\sigma}{\xi_{i}f_{\sigma}d\sigma}=\frac{1}{2\Delta V}\int_{\sigma}\left|\xi_{i}\right|\frac{\left|\xi_{i}\right|}{\left|\xi_{i}\right|} \Delta x_{i}\frac{\partial f}{\partial x_{i}}d\sigma
=\frac{1}{2\Delta V}\int_{\sigma}\tau\xi_{i}^{2} \frac{\partial f}{\partial x_{i}}d\sigma
\label{balance2a}
\end{dmath}

and using the Gauss-Ostrogradsky formula it is possible to transform Eq.~(\ref{balance2a}) to the differential form:
\begin{dmath}
\frac{\partial f}{\partial t}+\nabla\cdot\left(\boldsymbol\xi f^{j}_{M} \right)=\frac{\tau}{2}\frac{\partial}{\partial x_{i}}\frac{\partial}{\partial x_{k}}\xi_i\xi_k  f^{j}_{M}
\label{balance1}
\end{dmath}

Here the quasi magnetohydrodynamics system of equation involves explicitly two $\tau$ parameters. Hydrodynamics processes are introduced by the quantity $\tau$ that corresponds to the time of free distance flight of particles, or the characteristic time of particle collisions. By analogy the quantity $\tau_m$ is introduced as the characteristic time of propagation of magnetohydrodynamics by electromagnetic processes. 
The characteristic time values $\tau$ and $\tau_{m}$ are defined respectively for hydrodynamics and electromagnetic processes:
\begin{dmath}[compact]
\tau=\alpha\frac{\Delta x_{i}}{\bar{c}_{h}}\qquad\tau_{m}=\alpha_{m}\frac{\Delta x_{i}}{\bar{c}_{m}}
\end{dmath}

\hspace*{1.cm} where:\\
\hspace*{1.cm}$\Delta x_{n}$ is the size of the computational cell,\\
\hspace*{1.cm} $c_{h}$, $c_m$ are the sound speed and Alphen speed in the computational cell.

\vspace{0.25cm}
The introduction of the physical meaning of characteristic times $\tau$ and $\tau_m$ provides an important contribution to the understanding of the processes and the simplification of the numerical scheme. 

The evolution equations for the gas dynamics parameters and for the magnetic field are obtained from Eq.~(\ref{balance1}) by integration with the summation invariants $\phi\left(\boldsymbol\xi\right)=m,m\boldsymbol\xi,\frac{1}{2}m\boldsymbol\xi^{2},m\xi^{\ast}$ over the molecular velocities, under the assumption:
\begin{dmath}
\int{f^{j+1}\phi\left(\boldsymbol\xi\right)d\boldsymbol\xi}=\int{f^{j+1}_{M}\phi\left(\boldsymbol\xi\right)d\boldsymbol\xi}
\end{dmath}
The integration is performed as in Eq.~(\ref{mhd3d}) and Eq.~(\ref{MHDscheme}). The gas dynamics and magnetic field quantities are obtained respectively as the real and imaginary path of the integral in the complex space.

The compact form of the kinetic quasi magnetohydrodynamics system of equations can be written as:
\begin{dgroup}[noalign]
\begin{dmath}[compact]
\frac{\rho^{j+1}-\rho^{j}}{\Delta t}+\frac{\partial}{\partial x_{i}}\rho u_{i}=\frac{\partial w_{i}}{\partial x_{i}}
\end{dmath}
\begin{dmath}[compact]
\frac{\rho^{j+1}u_{i}^{j+1}-\rho^{j}u_{i}^{j}}{\Delta t}+\frac{\partial}{\partial x_{k}}\Pi_{ik}=\frac{\partial }{\partial x_{k}}\Pi_{ik}^{D}+\frac {\partial}{\partial x_{k}}w_{i}u_{k}
\end{dmath}
\begin{dmath}[compact]
\frac{E^{j+1}-E^{j}}{\Delta t}+\frac{\partial F_{i}}{\partial x_{i}}=\frac{\partial Q_{i}}{\partial x_{i}}+\frac{\partial}{\partial x_{i}}\Pi^{D}_{ik}u_{k}+\frac{\partial}{\partial x_{i}}\left(\frac{E+p}{\rho}+\frac{B^{2}}{2\rho} \right)w_{i}
\end{dmath}
\begin{dmath}[compact]
\frac{B_{i}^{j+1}-B_{i}^{j}}{\Delta t}+\frac{\partial }{\partial x_{k}} M^{B}_{ik}=\frac{\partial}{\partial x_{k}}\Pi_{ik}^{DB}
\end{dmath}
\label{qgde1}
\end{dgroup}

The left-hand part of the system of Eq.~(\ref{qgde1}) corresponds to the Euler system of equations. 
The right-hand of the kinetic quasi MHD Eq.~(\ref{qgde1}) contains dissipative terms. 
In comparison with other methods, the dissipative terms are obtained not by phenomenology with some assumption about magnetohydrodynamics processes but in consistency with the difference scheme of the Boltzmann equation. 

$\Pi_{ik}$ is the momentum flux density tensor for a perfect gas in magnetic field:
\begin{dmath}
\Pi_{ik}=\left(p+\frac{B^2}{2}\right)\delta_{ik}+\rho u_{i}u_{k}-B_{i}B_{k}
\end{dmath} 

$F_{i}$ is the heat transfer flux of a perfect gas in magnetic field:
\begin{dmath}[compact]
F_{i}=\left[\left(E+p+\frac{B^{2}}{2}\right)u_{i}-B_{i}u_{k}B_{k}\right]
\end{dmath}

$M^{B}_{ik}$ is the asymmetric product between velocity $\mathbf{u}$ and magnetic field flux $\mathbf{B}$:
\begin{dmath}[compact]
M^{B}_{ik}=u_{k}B_{i}-u_{i}B_{k}
\end{dmath}

The right hand part of the system of Eq.~(\ref{qgde1}) includes the dissipative terms:
\begin{dgroup}[noalign]
\begin{dmath}[compact]
w_{i}=\frac{\tau}{2}\frac{\partial}{\partial x_{k}}\left[\left(p+\frac{B^{2}}{2} \right)\delta_{ik}+\rho u_{i}u_{k}-B_{i}B_{k} \right]=\frac{\tau}{2}\frac{\partial}{\partial x_{k}}\Pi_{ik}
\end{dmath}

\begin{dmath}[compact]
\Pi^{D}_{ik}=\frac{\tau}{2}\left[p\frac{\partial u_{i}}{\partial x_{k}}+p\frac{\partial u_{k}}{\partial x_{i}}-\frac{2}{3}p\frac{\partial u_{m}}{\partial x_{m}}\delta_{ik}\right] \\
+\frac{\tau}{2}\left[\left(\frac{B^{2}}{2}\delta_{mk}-B_{m}B_{k}\right)\frac{\partial u_{i}}{\partial x_{m}}+\left(\frac{B^{2}}{2}\delta_{im}-B_{i}B_{m}\right)\frac{\partial u_{k}}{\partial x_{m}}-\left(\frac{B^{2}}{2}\delta_{ik} -B_{i}B_{k}\right)\frac{\partial u_{m}}{\partial x_{m}}\right] \\
+\frac{\tau}{2}\left[B_{m}\left(-B_{k}\frac{\partial u_{i}}{\partial x_{m}}-B_{i}\frac{\partial u_{k}}{\partial x_{m}}+B_{n}\frac{\partial u_{n}}{\partial x_{m}}\delta_{ik}\right)\right]
+\frac{\tau}{2}\left[\rho u_{i}u_{m}\frac{\partial u_{k}}{\partial x_{m}}+ u_{i}\frac{\partial p}{\partial x_{k}}+ u_{i}\frac{\partial}{\partial x_{k}}\frac{B^{2}}{2}- u_{i}\frac{\partial}{\partial x_{m}}B_{m}B_{k}\right] \\
+\frac{\tau}{2}\left[u_{m}\frac{\partial p}{\partial x_{m}}+\gamma p \frac{\partial u_{m}}{\partial x_{m}} \right]\delta_{ik}
+\frac{\tau}{2}\left[B_{n}^{2}\frac{\partial u_{m}}{\partial x_{m}}-B_{n}B_{m}\frac{\partial u_{n}}{\partial x_{m}}+B_{n}u_{m}\frac{\partial B_{n}}{\partial x_{m}}\right]\delta_{ik}\\
+\frac{\tau}{2}\left[-B_{i}B_{k}\frac{\partial u_{m}}{\partial x_{m}}+B_{i}B_{m}\frac{\partial u_{k}}{\partial x_{m}}-B_{i}u_{m}\frac{\partial B_{k}}{\partial x_{m}}\right] 
+\frac{\tau}{2}\left[-B_{k}B_{i}\frac{\partial u_{m}}{\partial x_{m}}+B_{k}B_{m}\frac{\partial u_{i}}{\partial x_{m}}-
B_{k}u_{m}\frac{\partial B_{i}}{\partial x_{m}}\right]
\end{dmath}

\begin{dmath}[compact]
Q_{i}^{D}=\frac{\tau}{2}\left[\frac{5}{2}p\frac{\partial}{\partial x_{i}}\frac{p}{\rho}\right]\\
+\frac{\tau}{2}\left[\frac{5}{2}\left(\frac{B^{2}}{2}\delta_{ik}-B_{i}B_{k}\right)\frac{\partial}{\partial x_{k}}\frac{p}{\rho}\right]\\
+\frac{\tau}{2}\left[\frac{3}{2}\left(p\delta_{ik}+\frac{B^{2}}{2}\delta_{ik}-B_{i}B_{k}\right)\frac{\partial}{\partial x_{k}}\frac{B^{2}}{2\rho}
-\left(p+\frac{B^{2}}{2} \right)\frac{\partial}{\partial x_{k}}\frac{B_{i}B_{k}}{\rho} 
-\frac{B_{i}B_{k}}{\rho} \frac{\partial}{\partial x_{k}}\frac{B^{2}}{2}\right]\\
+\frac{\tau}{2}\left[\rho u_{i}u_{k}\frac{\partial}{\partial x_{k}}\frac{3}{2}\frac{p}{\rho} \right]
+\frac{\tau}{2}\left[\rho u_{i}u_{k}\left(p+B^{2}\right)\frac{\partial}{\partial x_{k}}\frac{1}{\rho}
-u_{i}B^{2}\frac{\partial u_{k}}{\partial x_{k}}\right]
+\frac{\tau}{2}\left[u_{i}B_{m}\left(B_{m}\frac{\partial u_{k}}{\partial x_{k}}-B_{k}\frac{\partial u_{m}}{\partial x_{k}} +u_{k}\frac{\partial B_{m}}{\partial x_{k}}\right)\right]\\
+\frac{\tau}{2}\left[\frac{1}{2}\rho u_{i}u_{k}\left(\frac{B^{2}}{2}\frac{\partial}{\partial x_{k}}\frac{1}{\rho}-\frac{1}{\rho}\frac{\partial}{\partial x_{k}}\frac{B^{2}}{2} \right) \right]
+\frac{\tau}{2}\left[B_{i}B_{m}\left(-u_{k}\frac{\partial u_{m}}{\partial x_{k}}-\frac{1}{\rho}\frac{\partial p}{\partial x_{m}}-\frac{1}{\rho}\frac{\partial}{\partial x_{m}} \frac{B^{2}}{2}+\frac{1}{\rho}\frac{\partial}{\partial x_{k}}B_{m}B_{k} \right) \right]
\end{dmath}

\begin{dmath}[compact]
\Pi_{ik}^{DB}=\frac{\tau_{m}}{2}\left[\frac{1}{\rho}\left(p+\frac{B^{2}}{2} \right)\left(\frac{\partial B_{i}}{\partial x_{k}}-\frac{\partial B_{k}}{\partial x_{i}}\right) \right]\\
+\frac{\tau_{m}}{2}\left[\left(p+\frac{B^{2}}{2} \right)\left( B_{i}\frac{\partial}{\partial x_{k}}\frac{1}{\rho}-B_{k} \frac{\partial}{\partial x_{i}}\frac{1}{\rho}\right) \right] 
+\frac{\tau_{m}}{2}\left[ u_{k}B_{m}\frac{\partial u_{i}}{\partial x_{m}} -u_{i}B_{m}\frac{\partial u_{k}}{\partial x_{m}}    \right]\\
+\frac{\tau_{m}}{2}\left[ \frac{1}{\rho}B_{i}B_{m}\frac{\partial B_{k}}{\partial x_{m}} -\frac{1}{\rho}B_{k}B_{m}\frac{\partial B_{i}}{\partial x_{m}}    \right]
+\frac{\tau_m}{2}\left[u_{k}B_{i}\frac{\partial u_{m}}{\partial x_{m}}-u_{k}B_{m}\frac{\partial u_{i}}{\partial x_{m}}+u_{k}u_{m}\frac{\partial B_{i}}{\partial x_{m}}\right]
+\frac{\tau_m}{2}\left[B_{i}u_{m}\frac{\partial u_{k}}{\partial x_{m}}+\frac{B_{i}}{\rho}\frac{\partial p}{\partial x_{k}}+\frac{B_{i}}{\rho}\frac{\partial }{\partial x_{k}}\frac{B^{2}}{2}-\frac{B_{i}}{\rho}\frac{\partial}{\partial x_{m}}B_{k}B_{m}\right] \\
+\frac{\tau_m}{2}\left[-u_{i}B_{k}\frac{\partial u_{m}}{\partial x_{m}}+u_{i}B_{m}\frac{\partial u_{k}}{\partial x_{m}}-u_{i}u_{m}\frac{\partial B_{k}}{\partial x_{m}}\right]
+\frac{\tau_m}{2}\left[-B_{k}u_{m}\frac{\partial u_{i}}{\partial x_{m}}-\frac{B_{k}}{\rho}\frac{\partial p}{\partial x_{i}}-\frac{B_{k}}{\rho}\frac{\partial }{\partial x_{i}}\frac{B^{2}}{2}+\frac{B_{k}}{\rho}\frac{\partial}{\partial x_{m}}B_{i}B_{m}
\right]
\end{dmath}
\label{dissipative}
\end{dgroup}

The dissipative terms appear because the construction of the quasi magnetohydrodynamics system is based on the assumption that the distribution function slightly changes over the distance between neighborhood cells, what is related to the characteristic times $\tau$ and $\tau_m$. It was shown in \cite{Chetver_2} that the dissipative terms of the quasi gas dynamics system are small in comparison with the convective terms with the condition of cell size equivalent to the free path they converge to the viscous terms of the corresponding Navier-Stokes equations. The corresponding dissipative terms are associated with real physics processes. An important remark is that in this case the gas dynamics parameters such as viscosity and heat conductivity are obtained from the kinetic theory.

The Navier-Stokes viscosity is identified as the first term of  Eq.~(\ref{dissipative}b):
\begin{dmath}[compact]
\Pi^{NS}_{ik}=\frac{\tau}{2}\left[ p\frac{\partial u_{i}}{\partial x_{k}} 
+p\frac{\partial u_{k}}{\partial x_{i}}-\frac{2}{3}p\frac{\partial u_{m}}
{\partial x_{m}}\delta_{ik} \right ]
=\mu\left[\frac{\partial u_{i}}{\partial x_{k}}+\frac{\partial u_{k}}{\partial x_{i}}
-\frac{2}{3}\frac{\partial u_{m}}{\partial x_{m}}\delta_{ik}\right]
\end{dmath}
where the bulk viscosity component is neglected and the shear viscosity $\mu$ is related to the gas pressure $p$ and the characteristic time $\tau$ as $\mu=\frac{\tau}{2} p$. 

The Navier-stokes thermal flux vector is identified as the first term of Eq.~(\ref{dissipative}d):
\begin{dmath}[compact]
q_{i}=\frac{\tau}{2}\left[\frac{5}{2}p\frac{\partial}{\partial x_{i}}\frac{p}{\rho}\right]
=k\frac{\partial T}{\partial x_{i}}
\end{dmath}
with $T$ gas temperature and $k$ thermal coefficient expressed as $k=\frac{1}{Pr}\frac{5}{2}R\frac{\tau}{2} p$, 
with $Pr$ Prandl number.

A similar analysis of the dissipative terms of the electromagnetic processes gives the estimation of their smallness.  With correct conditions for the size of cells the equation converges to the correct representation of magnetic viscosity. The gas resistivity is identified as the first 
term of Eq.~(\ref{dissipative}c) and also appeares as a result of the kinetic formulation:
\begin{dmath}[compact]
\Pi^{B}_{ik}=\frac{\tau_{m}}{2}\left[\left ( p+\frac{B^2}{2} \right)
\left(\frac{\partial B_{i}}{\partial x_{k}}-\frac{\partial B_{k}}{\partial x_{i}}\right)\right]
=\eta \left[\frac{\partial B_{i}}{\partial x_{k}}-\frac{\partial B_{k}}{\partial x_{i}}\right]
\end{dmath}
with the resistivity $\eta = \frac{\tau_m}{2}\left( p+\frac{B^2}{2}\right ) $.

\section{Computational Algorithm}
The computational algorithm uses a Cartensian, staggered, divergence free mesh configuration. A detailed description is presented in  ~\cite{Balsara_1, Balsara_2}, in order to preserve the condition $\nabla\cdot \mathbf{B}=0$.

\begin{figure}[htp]
	\centering
	\includegraphics[height=0.5\textwidth]{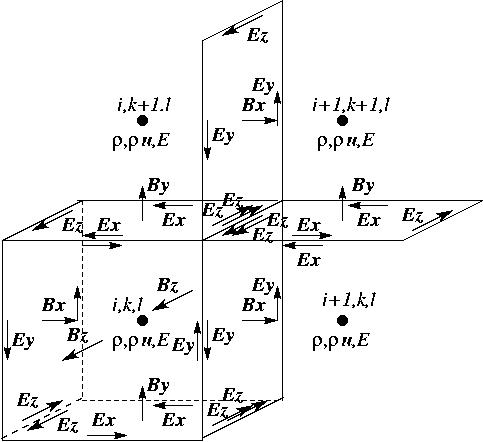} 
	\caption{3D Computational Domain}
	\label{fig:3D_domain}
\end{figure}

Fig.~\ref{fig:3D_domain} shows the four neighbour to the cell ($i,k,l$) used in evaluations of the hydrodynamics and electromagnetic variables. 

The hydrodynamics observables - mass density, momentum and energy density are defined at the cell center. The components of the magnetic field are defined at the face centers of the cells. A duality is established between the electric field and the fluxes. This duality is utilized to obtain the electric field at the edges of the computational cell through a reconstruction process that is applied directly to the properly upwinded fluxes. The electric field is then utilized to make an update of the magnetic fields that preserves the solenoidal nature of the magnetic field and ensures that the magnetic field in a magnetohydrodynamics model remains strictly solenoidal up to discretization errors.

Generally the explicit numerical scheme is used model, considering that it is perspective for the modern high performance computing systems  due to the logical simplicity and efficiency of the algorithms.

The finite volume method is used to update the conserved observables, mass, momentum and energy, by calculating the fluxes of this observables across the cell face. 
Updating the magnetic field is a more complicated procedure and is performed via electric field integration along the edge of the cells, as showh on Fig.~\ref{fig:3D_domain}.   
The distribution function method is proposed in the calculation as described above. 

The explicit scheme is used in the time evolution for the integration of the quasi magnetohydrodynamics system of equations.
The code uses a variable time step. The time step in an explicit scheme is controlled by a Courant type conditions on the time step estimation ~\cite{Chetver_2}.  

\section{Results of Numerical Modeling}
The computational framework is created on the basis of Fortran 90 and c++, with parallel implementation on MPI.

The demonstration of the performance of the method is performed on the basis of the solution of the spherical expansion problem of ionised gas and the solution of the expansion of an ionised gas in strong magnetic field.
  
The simulations are performed for a Cartesian rectangular mesh $100\times 100\times 100$ in the physics domain [0,1] . 
 
The initial conditions consist of a sphere with radius 0.1 placed in the center of the physical region with pressure of 100 in comparison to the overall represented area with pressure 1. For the study of ionized gas in a strong magnetic field the uniform magnetic field aligned with the  $z$ coordinate is added to the initial conditions.    

\begin{figure}[htp]
\centering
	\includegraphics[height=0.5\textwidth]{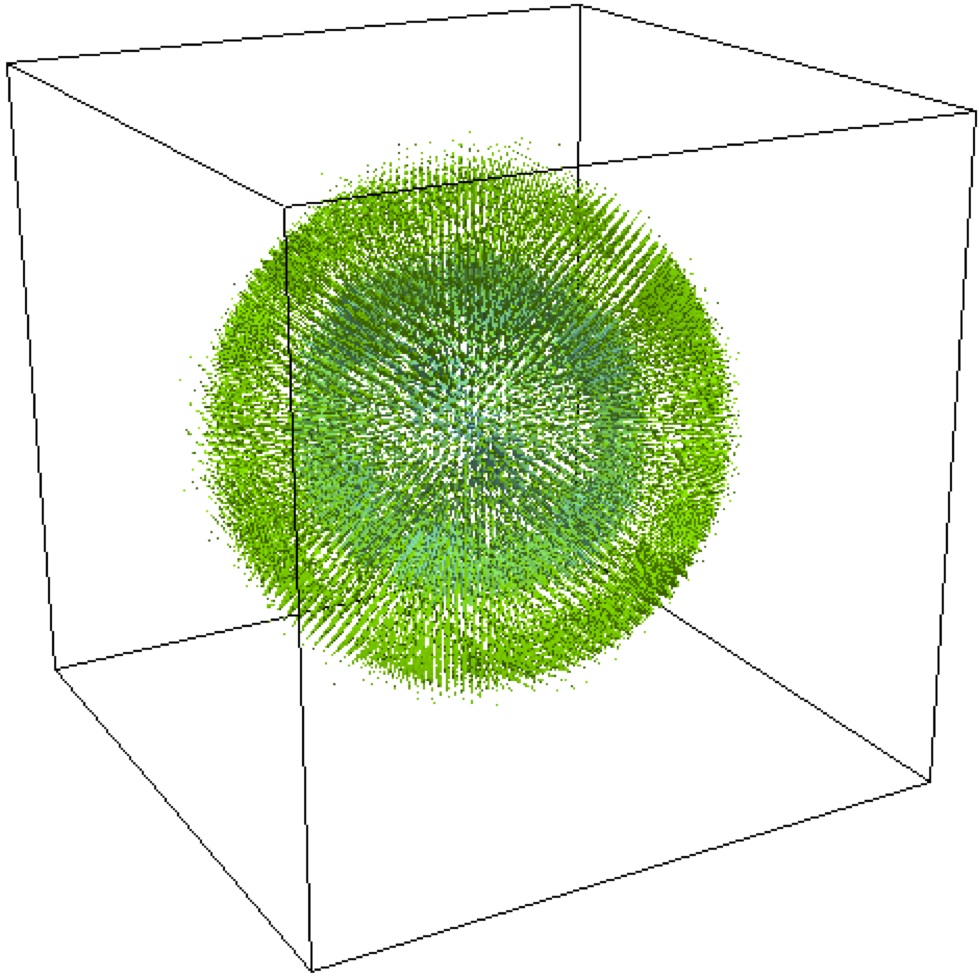}
	\caption{$3D$ view of the gas expansion without magnetic field}
	\label{fig:3D_exp}
\vspace*{0.5cm}
	\includegraphics[height=0.5\textwidth]{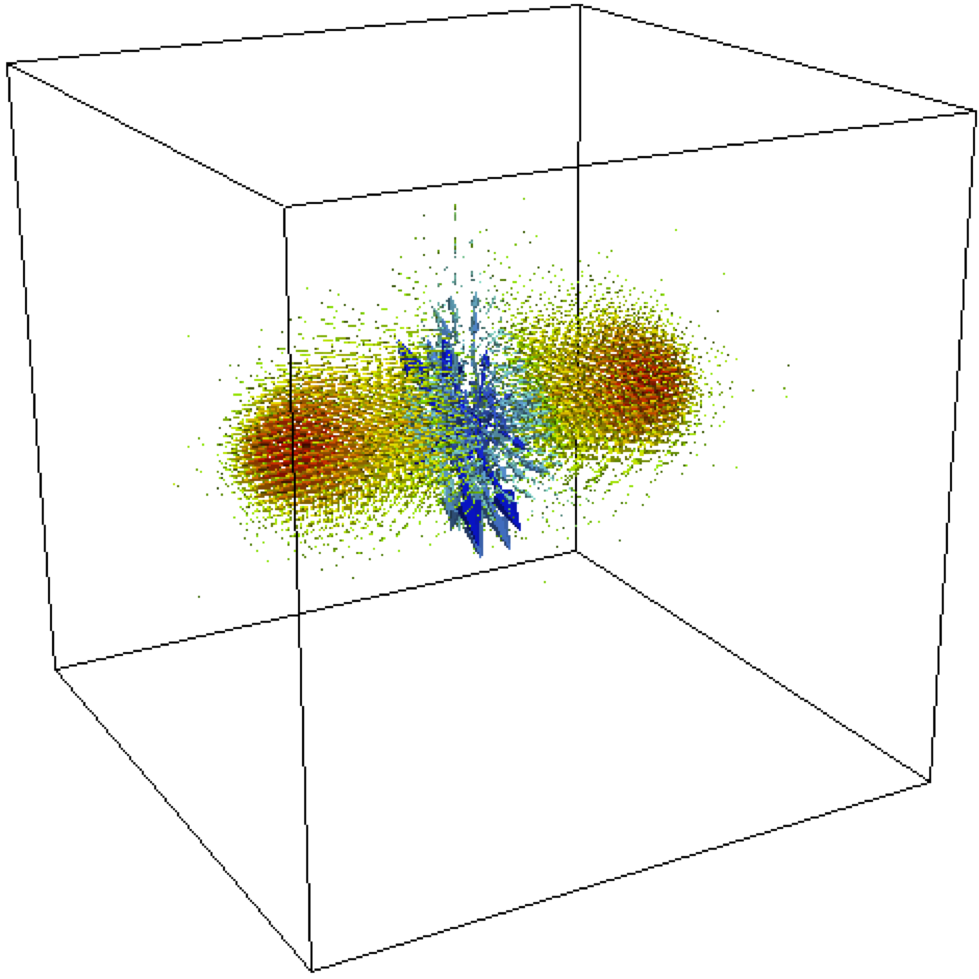}
	\caption{$3D$ view of the conductive gas expancion in strong magnetic field}
	\label{fig:3D_B_exp}
\end{figure}

Fig.~\ref{fig:3D_exp},\ref{fig:3D_B_exp} present the state of the 3D simulation of the processes for relative time 0.03. 
On the 3D pictures the arrows represent the velocities of the ionised gas and the color represents the density of gas. 3D figures clearly show the confinement of the ionized gas in the cylindrical area along $z$ due to the magnetic field.  

\begin{figure}[htp]
\centering
	\includegraphics[height=0.35\textwidth]{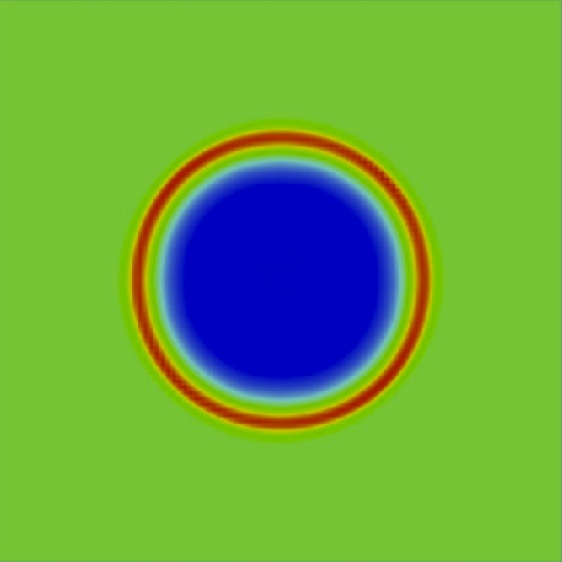}
	\includegraphics[height=0.35\textwidth]{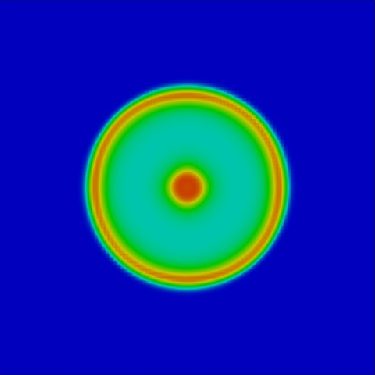}
	\caption{$2D$ gas density and 2D gas pressure projections}
	\label{fig:2D_proj_1}
\vspace{0.5cm}
	\includegraphics[height=0.35\textwidth]{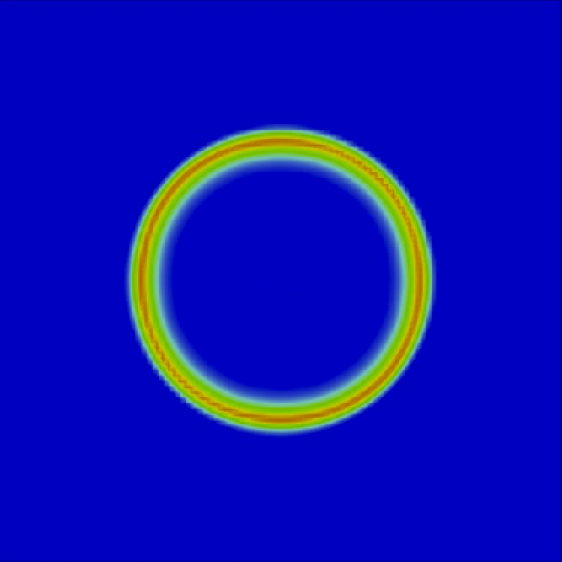}
	\includegraphics[height=0.35\textwidth]{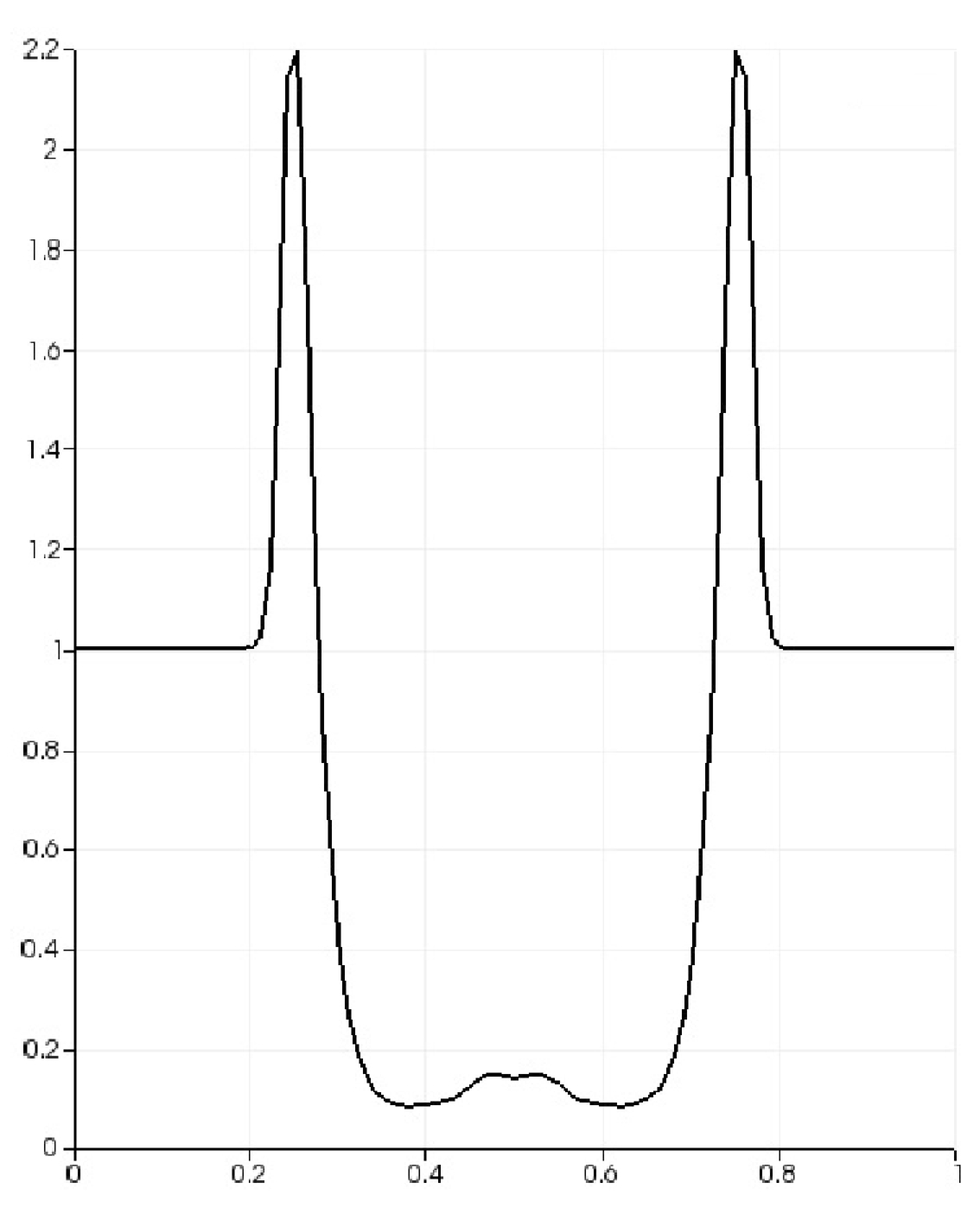}
\hspace*{0.65 cm}
	\caption{$2D$ kinetic energy and 1D density profile}
	\label{fig:2D_proj_2}
\end{figure}

Fig.~\ref{fig:2D_proj_1},\ref{fig:2D_proj_2} represent the $2D$ projections of the density, pressure and kinetic energy of the gas expansion without magnetic field and the $1D$ density profile for these condition.   

\begin{figure}[htp]
\centering
	\includegraphics[height=0.35\textwidth]{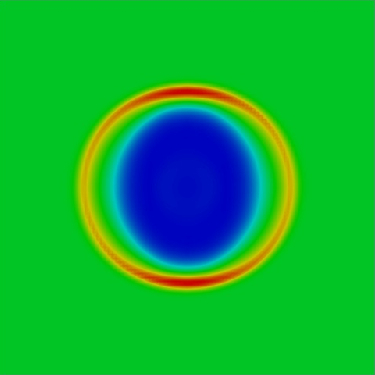}
	\includegraphics[height=0.35\textwidth]{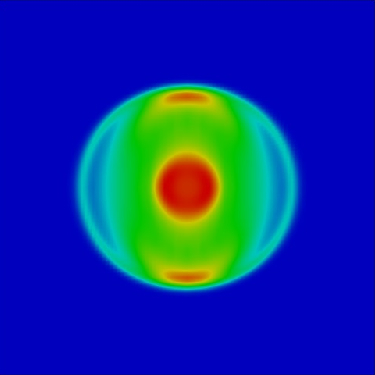}
	\caption{$2D$ gas density and $2D$ gas pressure projections in the magnetic field $5/\sqrt{\pi}$}
	\label{fig:2D_B_proj_1}
\vspace{0.5cm}
	\includegraphics[height=0.35\textwidth]{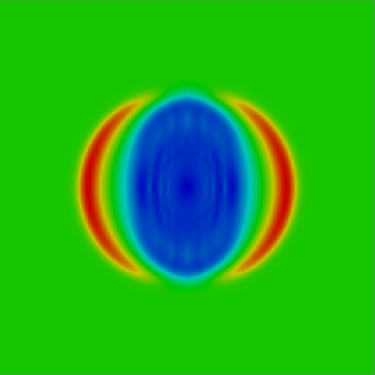}
	\includegraphics[height=0.35\textwidth]{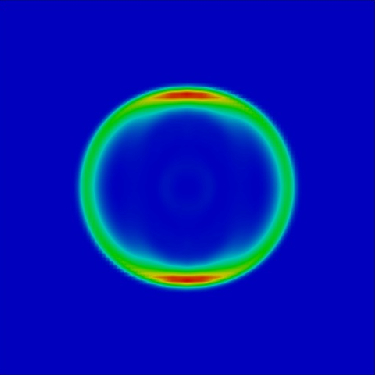}
	\caption{$2D$ magnetic pressure and $2D$  kinetic energy in the magnetic field $5/\sqrt{\pi}$}
	\label{fig:2D_B_proj_2}
	\includegraphics[height=0.35\textwidth]{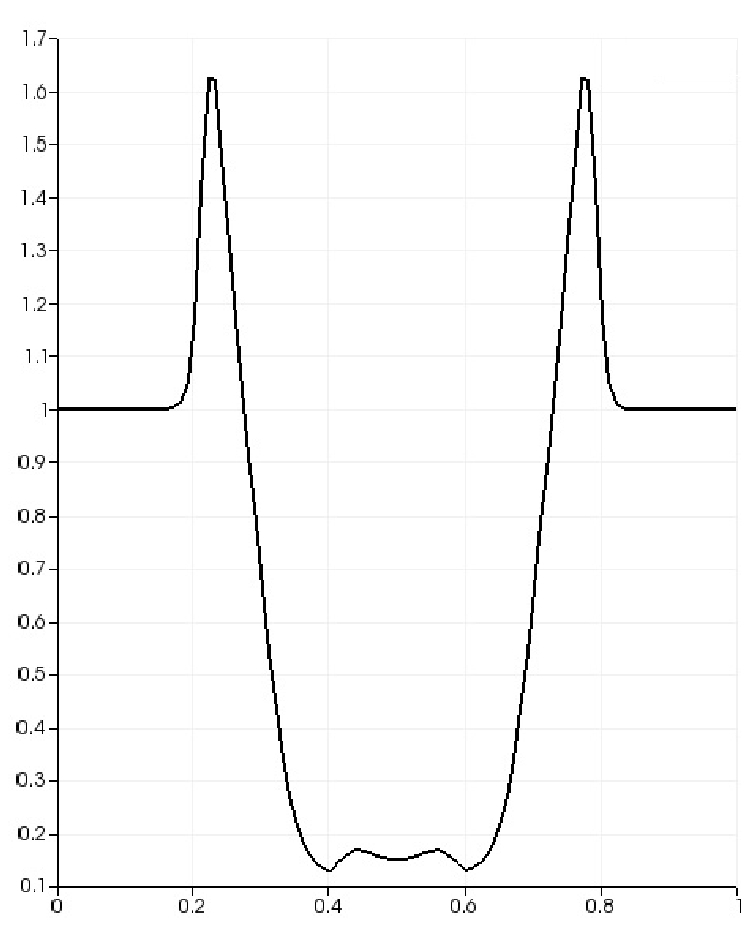}
	\caption{$1D$ density profile in the magnetic field $5/\sqrt{\pi}$   }
	\label{fig:1D_B_proj_1}
\end{figure}

Fig.~\ref{fig:2D_B_proj_1},\ref{fig:2D_B_proj_2} represents the $2D$ projections of the density, pressure, magnetic pressure and kinetic energy for the gas expansion problem of the ionized gas  with initial magnetic field of $5/\sqrt{\pi}$ and Fig.~\ref{fig:1D_B_proj_1} shows the $1D$ profile of the density for these conditions.   

\begin{figure}[htp]
\centering
	\includegraphics[height=0.35\textwidth]{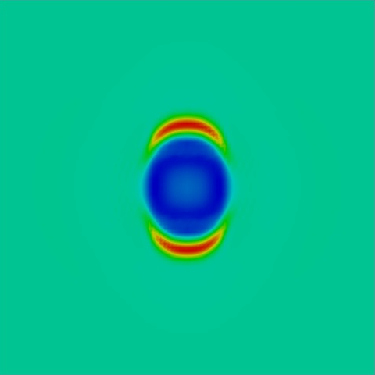}
	\includegraphics[height=0.35\textwidth]{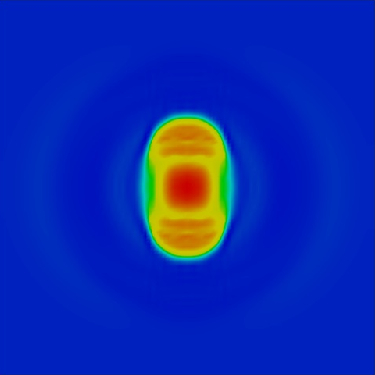}
	\caption{$2D$ gas density and 2D gas pressure projections in the magnetic field  $50/\sqrt{\pi}$     }
	\label{fig:2D_B1_proj_1}
\vspace{0.5cm}
	\includegraphics[height=0.35\textwidth]{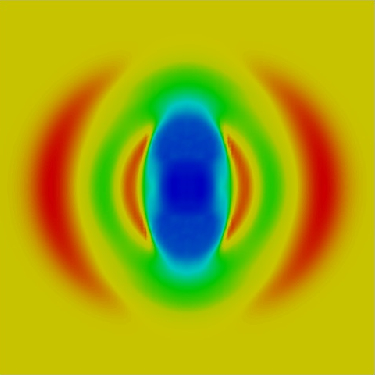}
	\includegraphics[height=0.35\textwidth]{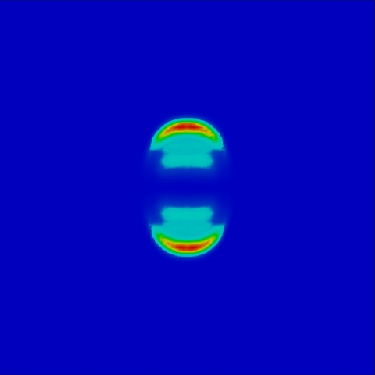}
	\caption{$2D$ magnetic pressure and $2D$ kinetic energy projections  in the magnetic field $50/\sqrt{\pi}$  }
	\label{fig:2D_B1_proj_2}
\vspace{0.5cm}
	\includegraphics[height=0.35\textwidth]{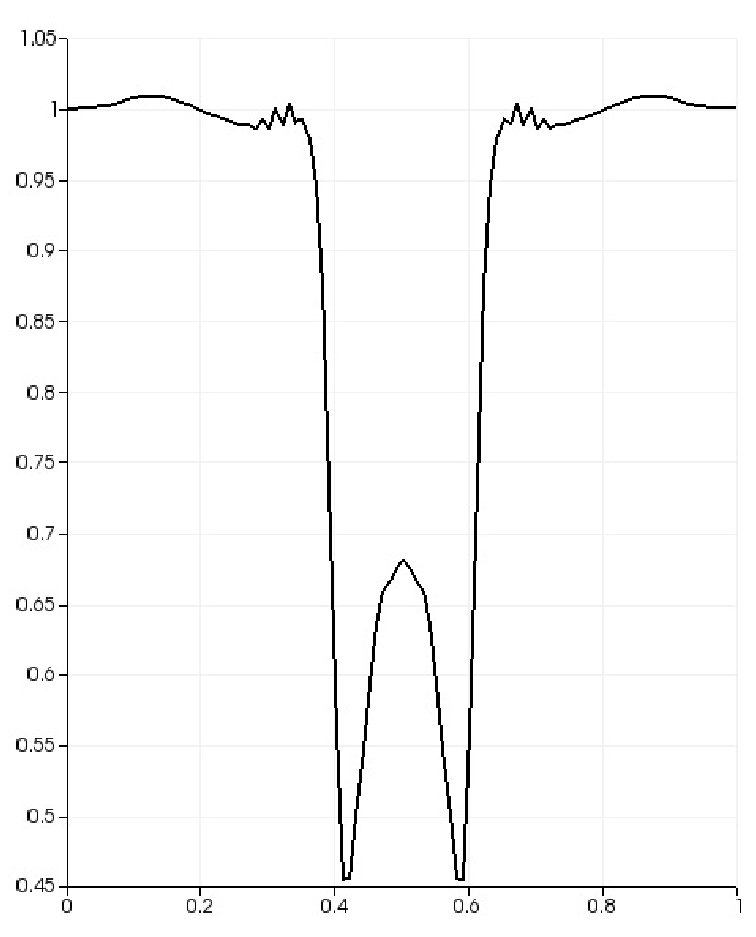}
	\caption{$1D$ projection of the density in the  magnetic field $50/\sqrt{\pi}$  }
	\label{fig:1D_B1_proj_1}
\end{figure}

Fig.~\ref{fig:2D_B1_proj_1},\ref{fig:2D_B1_proj_2} represent the $2D$ projections of the density, pressure, magnetic pressure and kinetic energy for the gas expansion of the ionized gas in the strong magnetic field of $50/\sqrt{\pi}$  at time 0.01 and Fig.~\ref{fig:1D_B1_proj_1} represents the 1D profile of the density for these conditions.   

Similar studies of the problem of spherical explosion including the conditions with magnetic field are presented in  ~\cite{Tang_1}. The comparison of the results shows a reasonable agreement and will be analysed further.    

\section{Conclusions}
A new 3D kinetic algorithm has been developed for the solution of the magnetohydrodynamics problems. The novel feature of the method is that the local complex Boltzmann-like distribution function incorporated most of the electromagnetic processes terms. The fluxes of mass, momentum and energy across the cell interface as well as the magnetic field are calculated by integrating a local complex Boltzmann-like distribution function over the velocity space. Thus by using this distribution function to calculate the mass, momentum and energy fluxes, most of the electromagnetic contributions are calculated directly, i.e. one does not have to solve the hydrodynamics and magnetic force components separately or differently.

A staggered, divergence free mesh configuration is used for the evaluation of the electromagnetic behaviour.

Numerical examples demonstrate that the proposed method can achieve high numerical accuracy and resolve strong shock waves of the magnetohydrodynamics problems.





\end{document}